# NEW SEQUENCE ALIGNMENT ALGORITHM USING AI RULES AND DYNAMIC SEEDS

Suchindra [1] and Preetam Nagaraj[2]

[1]Department of Engineering, National Institute of Mental Health and Neurosciences, Karnataka State Govt, Bangalore, India
Suchindra.Karnataka.gov.in@gmail.com

[2]Department of Engineering, IBM, Bangalore, India
Preetn12@in.ibm.com

## ABSTRACT

*DNA sequence alignment is important today as it is usually the first step in finding gene mutation, evolutionary similarities, protein structure, drug development and cancer treatment. Covid-19 is one recent example. There are many sequencing algorithms developed over the past decades but the sequence alignment using expert systems is quite new. To find DNA sequence alignment, dynamic programming was used initially. Later faster algorithms used small DNA sequence length of fixed size to find regions of similarity, and then build the final alignment using these regions. Such systems were not sensitive but were fast. To improve the sensitivity, we propose a new algorithm which is based on finding maximal matches between two sequences, find seeds between them, employ rules to find more seeds of varying length, and then employ a new stitching algorithm, and weighted seeds to solve the problem.*

## 1. INTRODUCTION

In computational biology or Bioinformatics, a sequence is either an RNA, DNA or a protein string made up of their representative character set. DNA ( A, C, G, T) , RNA ( A, C, G, U) and protein molecules (A, R, N, D, C, Q, E, G, H, I, L, K, M, F, P, S, T, W, Y, V) can be re represented as strings of letters from their alphabet set [1] [2] [3] [4] [5] [32] [33].

A sequence alignment is a way of arranging these sequences made of their representative characters with an objective to find the regions of '*similarity*'. These similarities would then provide additional information on the functional, structural, evolutionary, and other interest between the sequences in study. Aligned sequences are represented in rows, stacked up one on top of the other as shown below in Figure 1.

```
T T A T A G A G G _ A C A _ C G
    | | | |     | |   | | |   | |
_ _ A T A G _ G G G A C A T G G
```

Fig. 1. Example of a sequence alignment. [5]

In fig 1, there are regions, where the two sequences are aligned perfectly, these regions are called 'similar region. In some regions, special characters such as '-', also known as indels are present. These indels represents a mutation (change) or could be looked at as deletion from the other sequence's point of view [5].

Pairwise sequence alignment is used to find conserved regions in two sequences. Multiple sequence alignments are used to find common regions in more than 2 sequences. Pairwise Sequence alignment [5]. Pairwise sequence alignment is a first step and represents at times the first step in many bioinformatics solutions. In many multiple sequence alignment algorithms this remains the first step, especially the linear multiple sequence alignment algorithms.

Pairwise sequencing as in Fig 1, is alignment between 2 sequences of the same kind, DNA, RNA or Protein. The alignment would then throw some knowledge on the divergence of one sequence over the other in some cases or similarity in some cases.

Pairwise sequence alignment can be classified into local sequence and global sequence alignment. Local sequence alignment finds best approximate subsequence match within two sequences while the global sequence alignment takes the entire sequence into consideration [5].

Local sequence alignments are therefore designed to search subregions within the two sequences. For finding similar (biologically conserved) regions, which may or may not be preserved in order or orientation, local sequence alignment is very useful. It is typically used to find similarity between two divergent sequences and for fast database searches for similar sequences [5]. Since, it is trying to find subregions and not the sequence in its entirety, local sequence alignment usually takes less computation time when compared to global sequence alignment algorithms [5].

Some of the most popular local sequence algorithms are Smith-Waterman [6], FASTA [7], BLAST [8], GappedBLAST[9], BLASTZ[10], PatternHunter[11], YASS[12], LAMBDA[13], USearch [14], LAST [15], and ALLAlign [16].

Popular global sequence alignment algorithms including Optimal, and Heuristic based are AlignMe [17], Needleman and Wunsch [18], GLASS [19], WABA [20], AVID [21], and CHAOS [22]. We will not be discussing global alignment algorithms in this paper.

The alignment algorithms performances in the literature are based off several key measurements. This is very touchy, subjective topic and can vary from algorithm to algorithm. Some are based on type of sequence (they can be only for DNA, or RNA), length (some algorithms can be good for shorter than others), Measure of accuracy (since there is no standard here, this is a controversial and subjective), Speed of alignment (time taken to align the sequences in study) and lastly memory efficiency (how much memory is taken to find the alignment) [5]. Most of the algorithms do not require any Operating System writer-reader lock help in aligning the sequences [34].

## 2. LITERATURE

In this section we will talk about the popular algorithms in local sequence alignment. Our algorithm is greatly influenced by previous algorithms and has taken clues and has questioned at times on the approach and ideas of the previous algorithms. We first start the section with optimal and then heuristic algorithms.

Smith-Waterman Algorithm is an optimal local sequence alignment algorithm, employing a technique called '*Dynamic Programming*', where a problem is broken into smaller problems and solving these smaller problems recursively [6]. The solutions to the subproblems are saved and are brought together to find the solution to the entire problem. This optimal algorithm produces an optimal local sequence alignment between two sequences S1 and S2 of length m and n, respectively, in time and space equal to $O(mn)$ [6]. As the sequence length increase the performance decreases exponentially.

To overcome short comings of the optimal algorithm, Heuristic algorithms were developed later. Heuristic algorithms find a near-optimal solutions sacrificing little sensitivity for speed. Since it is near perfect and can calculate long sequences running in billions of characters, heuristic algorithms are preferred. All heuristic algorithms run in stages, stages where they find the maximum subsequence and try to find the regions arounds them to get the sequences local sequence alignment. These subsequences are called seeds or anchors depending on whether the subsequence is one large word (anchor) or small words made of few characters (seed).

The first heuristic algorithm we are going to talk is FASTA, which stands for FAST-ALL is a heuristic algorithm developed by Lipman and Pearson [7] [2]. FASTA uses a look-up table to find perfect subsequence matches of size 'l' and then hashing them. Time is saved for searching seeds of length 'l', then proceeds to find such seeds in a diagonal path, since the final alignment is more likely to be found in this diagonal length. It then uses a directed weighted graph for regions in between the seeds along the best diagonal found to stitch the seeds. The advantage of FASTA over Optimal algorithm is the speed but if there are 2 optimal diagonals or if the seeds are smaller than the size deployed in FASTA, then the algorithm loses considerable sensitivity [2].

Basic Local Alignment Search Tool BLAST [8], uses a look-up table to identify seeds and is faster than FASTA [2]. Sliding window technique is employed to find all good neighbors seeds for each seed it finds in both directions [2]. When all seeds are found, it then proceeds to find the seeds (HSP, high scoring pairs), and extend them until they fall under a threshold score 'k'. These HSP are then stitched using a restricted dynamic programming which is a version of Smith-Waterman Algorithm [6].

BLAT – BLAST like alignment tool [23] is much faster than BLAST. BLAT differs from previous algorithms in the way sequences are indexed. "non-overlapping seeds of S2 are run through the database of sequence and then a new scan is run linearly through the S1, whereas BLAST builds an index of S1 and then scans linearly through the database" [23]. This saves time. After this stage, it then searches for seeds with some mismatches 'n' in them around the seeds it found earlier. Then the HSPs of seeds and mismatch seeds are extended like BLAST to form a final alignment. As with BLAST, BLAT cannot find smaller homologous regions as the seeds taken as not small enough.

BLASTZ [8], is the fastest among the BLAST family of algorithms, employs a different method. All repeats in the sequence are removed [2]. It then looks for seeds of length 'l' with almost one-character transition. All seeds are then extended on both sides. For regions in between the seeds, it employs smaller seeds and uses optimal alignment to stitch these seeds to form final alignment. Since matched or repeat seeds are not used again, and transition seeds set to almost one, there is a possibility that this algorithm performs poorly when it is used for divergent sequences. Meaning, say a drosophila and a pig DNA.

PatternHunter [11] introduced a seed called spaced seed to further improve the sensitivity and speed. It uses a combination of priority queues variation of red-black tree, queue, and hash table to achieve speed [11]. A spaced seed is a generic seed which is converted from A, C, G, T to Binary form e.g.: 1010101,4 where 1 is a match and 4 is a score [5]. It then finds the best diagonal as in FASTA to find the final alignment [5]. The algorithm is written in JAVA, and encounters memory problems for long sequences.

UBlast [14] introduces a new technique by finding fewer good hits. Meaning, subsequences which are found least but are long, to improve speed on BLAST and MEGABLAST [24] which is algorithm from BLAST family. The technique is targeting more speed than sensitivity.

LAST [15] is recent algorithm. It uses adaptive alignment seeds; these adaptive seeds vary in length and the number of indels in them. So adaptive seeds can be of different lengths and weight. By weight, a score associated with the seed. The rest of the algorithm is very similar to BLAST. ALLAlign [16] is a new algorithm developed, however literature of this AWS based web algorithm is very limited.

LAMBDA [11] is new algorithm for protein sequence alignment. It implements a technique where there are more than 1 protein sequences as the target sequences to be aligned with a pre indexed database set of all other know sequences [2]. It is optimized for big or large biological data and uses a Suffix tree to get the maximal common subsequences or maximal unique sequences as our algorithm in this paper, amongst them and then goes about aligning these subsequences against a pre indexed database (pre indexed based off suffix array) [2].

MASAA [1] [3] introduced in 2008 is based on Ukkonen suffix [2] tree. The algorithm uses double indexing and back tracking and identifies maximum match subsequences (MMSS) [1][2][5]. In the subsequent stages, it finds perfect and near perfect seeds and stitches the local alignment in the last stages.

## 3. BASIC DEFINITIONS AND TERMINOLOGY

In this section, we introduce several fundamental definitions.

*Definition 3.1. Let 'L' be a set of characters called alphabet set. A sequence 'S' is an array of characters from L, such that they are all written contiguously from left to right and occupy a unique position in the sequence 'S'. If 'S' is a sequence, then |S| denotes the length of 'S' and S[i] denotes the i ℞ character of S [3].*

Example: Let 'S' is ACBCDB, here |S| =6 and S [3] =B.

Given two sequences, ATATAGAGGACACG and ATAGGGGACATGG, there are two long patters in these two sequences which are 'ATA' and 'GGACA' which make good pattern. We will define what this 'good' mean in the following sections. When the goal is to find common genes and genes which have mutated by a small percentage, it is necessary that we find the common patterns in the same region in both the sequences and compare their similarity.

Perhaps there could be some mutation or there could be all similar. Such pattern comparison makes more biological sense rather than patterns who different regions of the sequences are. Consider the sequences S1 and S2 as shown in Figure 1. In this example, we see that there are regions in S1 which are aligned with regions in S2. These common regions could be conserved regions that have not changed by evolution [5].

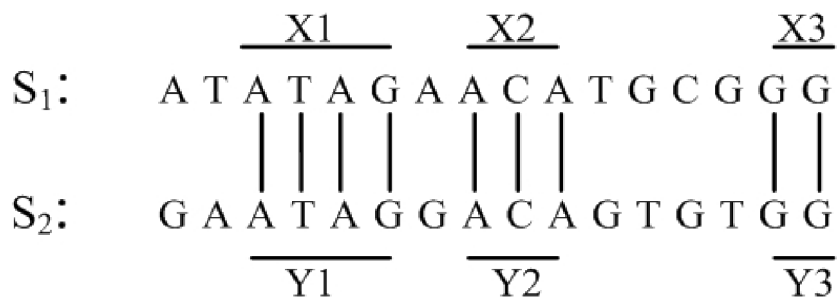

Figure 1: Conserved Regions in Two Sequences

In Figure 1, the conserved region X1 is identical to Y1. That is, every character in X1 matches the character in the corresponding position in Y1.

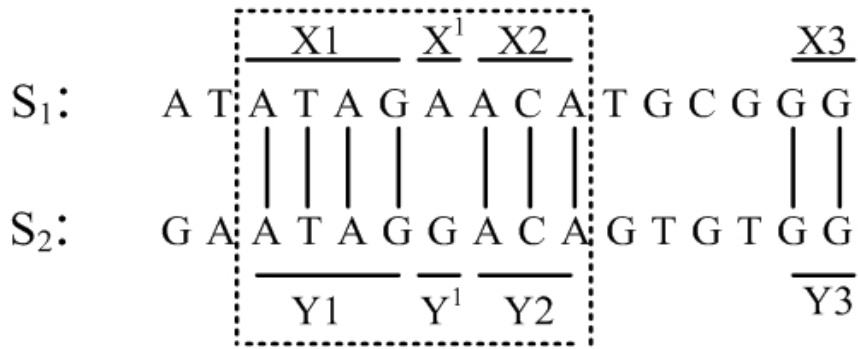

Figure 2: Highly Similar Regions in Two Sequences

Consider Figure 2 given below. From Figure 2, it is logical to consider that X1X′X2 and Y1Y′Y2 as conserved regions with a mutation at either X′ or Y′. The goal is to identify such highly similar regions [5], then they can be further checked in regions in between them to see if there are other similar conserved regions. Such pair of highly similar regions are referred to as 'seed' [23] [26] [25]. We formally define a seed below.

*Definition 3.2. X1 and X2 are considered as a seed if there exists a contiguous region X2 of m in S2, such that X1 is highly similar (≈) to X2 [2].*

High similarity between X1 and X2 could be

HS1: A perfect match, where every character in X1 matches the character in the corresponding position in Y1

HS2: é(X1, X2) ≥ t, for a given t and positions where base pairs (bps) must match (position) are not fixed.

HS3: X1 matches closely with X2, such that care (and do not care) positions are fixed unlike the other seeds already mentioned.

> HS3:1 K mismatches in do-not care positions and m-k matches in care positions (BLAT seeds) [23].
>
> HS3:2 K-mismatches in k do not care positions (Spaced seeds) [11].
>
> HS3:3 The score of é(X1, X2) in care position of $X_L, X_{L+1}...X_{L+P-1}$ and $Y_L, Y_{L+1}...Y_{L+\Theta 1}$ ≥ t, for a given t (Vector seeds) [25].

### 3.1. Performance Metric

Let S and T be database of sequences, T being the sequence from the database. The objective is to find an alignment of S and T, that has the maximum possible score for these two sequences [26]. From the literature, algorithms vary depending on one or more of the following metrics:

(1) Space and time efficient.

(2) Sensitive. The sensitivity is measured by four different scores, they are:

> (a) Percent identity score: representing the percent of the similar patterns that involves identical base pairs(bp) [27].

(b) Total column score (TCS): the number of similar patterns divided by the number of similar patterns in the reference sequence [28].

(c) Percent similarity number: Percent of pattern similarity that involves identical and similar matching residues. This applies to protein sequence alignments, where similar residues are amino acids that have similar physio-chemical properties [27].

## 4. SUFFIX TREE

In this section, we explain in detail the data structure used in algorithm. Ukkonen online suffix tree construction algorithm, [29] forms the first initial step of our algorithm. We begin this section by explaining the Ukkonen suffix tree building algorithm.

To aid the understanding of our proposed algorithm, we first present some terminology. Let S [0...N] be string indexed by the tree T. The leaf node corresponding to the $i^{th}$ suffix, S [i... N], is represented as $l$ ∈ An internal node, v, has an associated length L(v), which is the sum of edge lengths on the path from root to v. We represent by $ĝ(v)$, the string at v to represent the substring S [1...i + L(v)] where $l$ ∈ is any leaf under v. The suffix tree for an example, S = "MISSISSIPPI" is shown in Figure 3. The numbers at the bottom of leaf nodes represent the start of the suffix S [i.. N] that they represent [3]. The suffix tree is constructed incrementally by scanning the string from left to right, one character at a time. That is, suffix tree is built in m phases, one for each character. At the end of phase i, we will have tree $T_i$ which is the tree representing the prefix S [1...i].

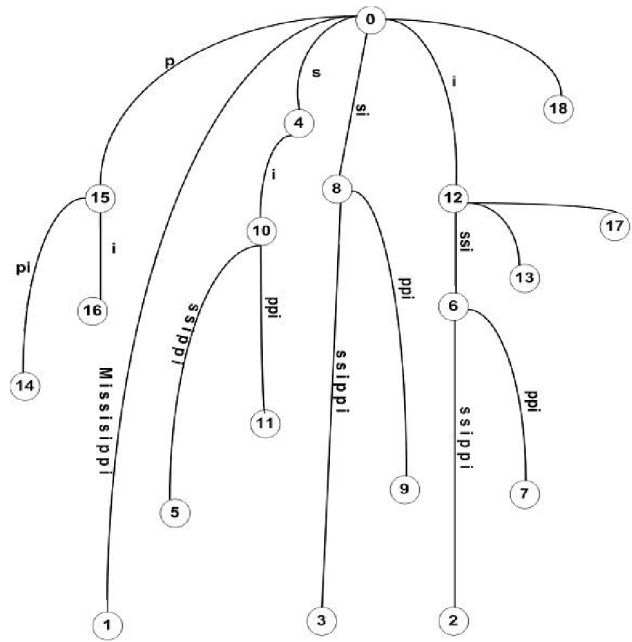

Figure 3: Suffix Tree for 'MISSISSIPPI'

In each phase i, we have i extensions, one for each character in the current prefix. At the end of extension j, we will have ensured that S[j..i] is in the tree $T_i$. There are four possible ways to extend S[j..i] with character i+1 [3].

(1) S[j..i] ends at a leaf. Add the character i+1 to the end of the leaf edge.

(2) There is a path through S[j..i], but no match for the i+1 character. Split the edge and create a new node, if necessary, then add a new leaf with character i+1.

(3) There is already a path through S[j..i+1].

(4) Do nothing.

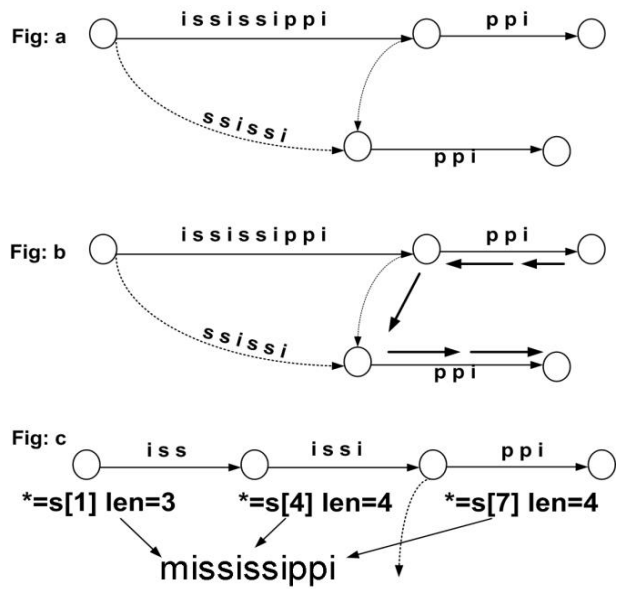

Figure 4: Speeding Up Steps to Build Suffix Tree

Definition 4.1. "Let a Ùdenote an arbitrary string, where a denotes a single character and Ù denotes a (possibly empty) substring. For an internal node v with path-label a Ù if there is another node(v) with path-label Ù then a pointer from v to s(v) is called a suffix link [3]. A suffix link sl(v) = w exists for every node v in the suffix tree such that if ê(v) = a Ù then ê(w) = Ù where a is a single character of the alphabet and Ùis a substring (possibly null) of the string. Note that sl(v) is defined for every node in the suffix tree. And, more importantly, sl(.) - the entire set of suffix links, forms a tree rooted at the root of T, with the depth of any node v in this sl(.) tree being L(v)" [30].

The suffix tree for, S = "MISSISSIPPI" with dashed edges between internal nodes representing suffix links is shown in Figure 5. To be fast and memory efficient, Ukkonen algorithm employs the following:

Step 1: The tree is supported with additional edges, called suffix links, that provide shortcuts to move across the tree quickly. These suffix links play a crucial role in reducing the running time of the algorithm [3].

Step 2: Skip/Count Trick as it is called: instead of stepping through each character, we know that we can just jump if the tree has common sub-strings. In other words, there are two branches having common sub-strings at different places, one can jump from one branch to the other branch [3], as shown in Figure 4.

Step 3: Edge-Label Compression, since we have a copy of the string, we do not need to store copies of sub-strings for each edge as shown in Figure 4.

Step 4: A match, if we find a match to our next character, we do not have to do anything as the sub-string now is already part of the built tree [3].

Step 5: Once a leaf, always a leaf. We do not need to update each leaf, since it will always be the end of the current string [30]

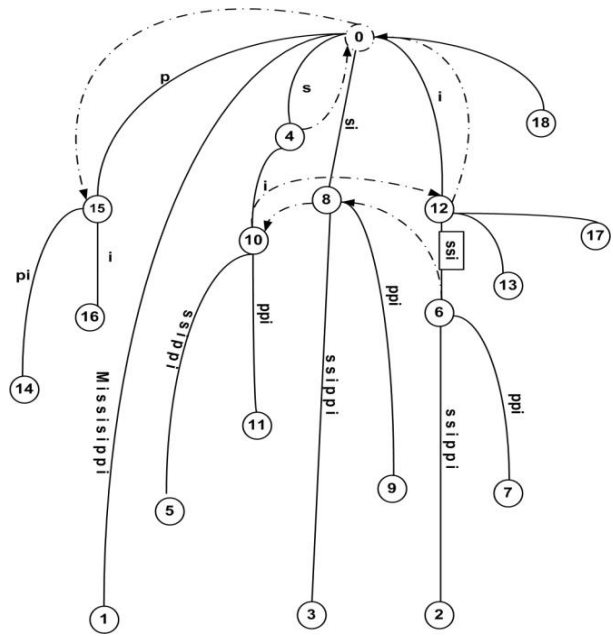

Figure 5: Suffix Links in Suffix Tree

## 5. ALGORITHM

Before we begin to describe the algorithm, we did like to describe some concepts and terminologies. A Maximum Match Sub-sequences is a region in DNA sequences which is well preserved through evolution and serves as a basis to our algorithm. The important criteria for pattern matching are to find these genes or conserved regions which make biological sense to the lab technicians. Hence quickly finding these patterns is of importance. We call such long conserved regions as Maximal Match Sub-sequence (MMSS) and therefore name our algorithm a Rule-Based Maximal Match Substring (RBMS) algorithm. There are regions in between these MMSS which are smaller in length which need to be identified too, these regions are called seeds, and both are shown in Figure 6.

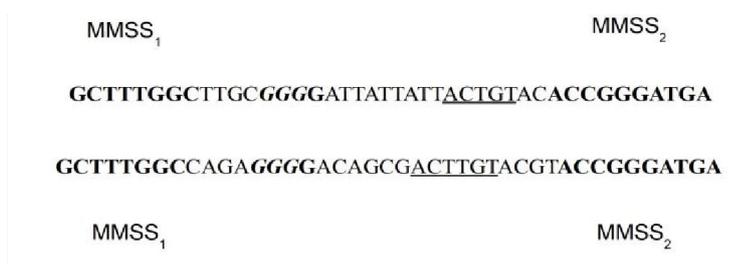

Figure 6: MMSS and seeds in between the MMSS Region

The algorithm identifies these MMSS with their position and number count associated with sequence/s and are stored in dynamic nest hash table of variable length. Dict = {ÁCGTA': {Ś1':{ '50' : 'Pos', '60':'Pos'}}}. When a new ACGTA is found in sequence S2, then it would add with S2 entry and so on. The algorithm would also make entries to these seeds in the DNA sequence starting from seed length 'l' and gradually coming down to 'l-n', These seed length are stored in the AI rules. These are taken in consideration only because of remove more noise and retain more quality in the pattern searched which make more biological recognition. AI rule is a simple if then statement, describing the length of the seed to be searched first and whether to merge the overlapping seeds or to discard. This makes the seeds more dynamic and can either increase the speed of the algorithm or decrease the sentivity.

## 5.1. General Algorithm Description

In this section, we will describe the algorithm. The algorithm is broken down to following steps or stages. The algorithm is shown below in Figure 7.

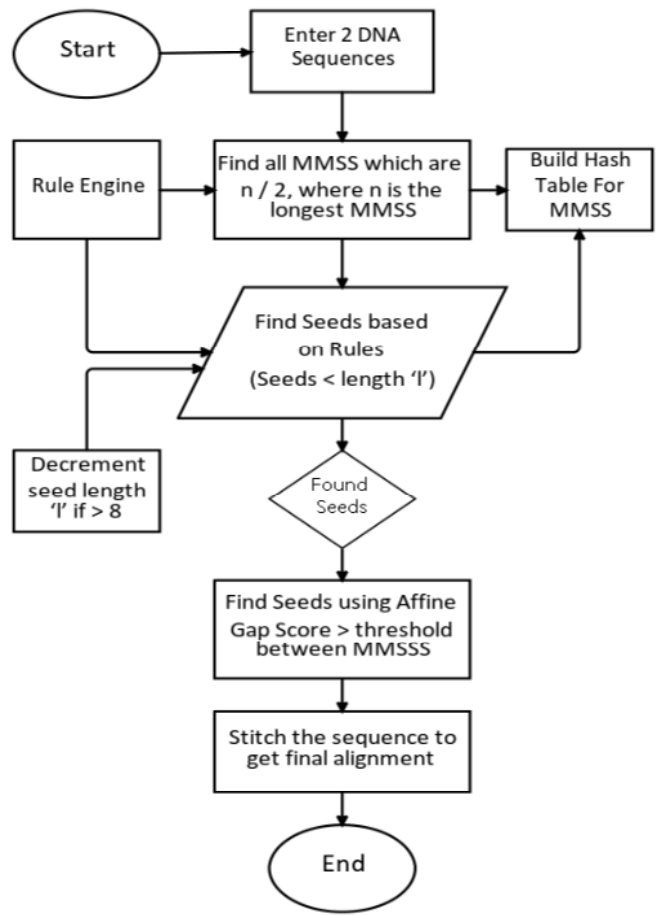

Figure 7: Algorithm Workflow

(1) Scan the DNA sequence, build the Suffix tree.

(2) Identify the MMSS and record them in a dynamic hash table or dictionary. In this step, the MMSS could only one but to get a maximal MMSS count, we identify all MMSS's which are more than half of the length of the longest are identified. The position and the sequences are identified and recorded in the dictionary. This is one of the rules in the rule engine. The rule engine is only a set of 'if then record', rules, which are enacted on the MMSS in this stage.

(3) In this stage, the seeds of length '*l*' are selected, and this is one of the variables we would like to research more in the future. The algorithm selects these seeds in the InterMMSS regions and record them in the same row as the MMSS record or augments it, this is crucial information and we do not want this information recorded as a separate row in the dictionary. Once identified, the algorithm would reduce the Seed length 'l' to k-1 unless k-1 is less than 8. Here there are different kind of rules are enacted.
   a. If there are overlapping MMSS, then both are joined to make one long seed
   b. Criss crossing MMSS are omitted and are sacrificed for speed.
   c. The weight of the seeds is given a score, using a affine gap score matrix, where is a penalty for mismatch
      i. A sperate score is also added depending on how far they are from MMSS.
(4) When there are no seeds found for a particular length 'l', then the algorithm moves to the selection stage. In this stage, only the top 10 percent of most MMSS are selected. This is a variable which can be changed. Finally, all the top frequent patterns are obtained.
(5) Stitching algorithm is then used to join, seeds with MMSS regions.
   a. The stitching algorithm would then combine the regions in between the seeds which are not necessarily perfectly aligned with the seeds, which are then stitched together with the MMSS. This step brings up the speed while sacrificing some sensitivity

## 5.2. Experimental Results and Discussion

In the experiments, we randomly generated sequences ranging from 100k to 500k. Although the algorithm is designed to scan multiple sequences in the database and then give the top 10 percent in a 1.3 million sequences. This algorithm could not be directly compared to any other similar algorithm on 1-1 comparison, without having the same exact number of sequences and same mixture of sequences in the database. Hence only pairwise sequences alignments are compared, and we check the run time with brute force (BF) method [5] and [24]. In brute force method, there is a character-to-character matching, using dynamic programming, which is very sensitive and is a gold standard for qualitative pattern matching algorithms, while BLASTZ is the modified version of BLAST which can find pattern at a faster speed. The comparison is shown in Table 1.

Table 1: Speed of the algorithms

| Length | RBMS | BLASTZ | Brute |
|---|---|---|---|
|  |  |  |  |
| 100000 | 11 | 11 | 987 |
| 200000 | 17 | 17 | 5660 |
| 300000 | 27 | 26 | 12499 |
| 400000 | 41 | 40 | 72895 |
| 500000 | 49 | 47 | 113563 |

To compare the quality of the pattern matching, we used a data set of homologous sequences which are about 10000 in number. The sequences are a mixture of genes from Drosophila to bacteria and fungus. We believe that for homologous sequences there was no point in comparing the brute force method as it would always return a higher number, therefore we compared it to BLASTZ algorithm.

Table 2: Algorithm Sensitivity for Homologous Sequences

| Exon | RBMS | BLASTZ |
|------|------|--------|
| 100% | 90   | 90     |
| 90%  | 95   | 95     |
| 70%  | 98   | 98     |
| 50%  | 100  | 100    |

Table 2 shows the percentage of exon coverage for all sequences in the dataset. The data set consists of more than 700 sequences, out of which 60 were selected. Both algorithms covered 100 exons, for 54 sequences. We see from the results that RBMS algorithm is almost on par with BLASTZ for homologous sequences, which we were not surprising, as MMSS identification from suffix tree and perfect matching in the inter MMSS region should have given it an advantage. On deeper analysis we think that, longer seeds taking precedence over smaller seeds in these regions could be the cause for this result. We see a better performance for divergent sequences.

Table 3 shows the percentage of Exon covered by different algorithms. We see that RBMS is performing better than BLASTZ. We believe that the BLASTZ sacrifices sensitivity for performance and there by does not look for more seeds in between the MMSS hence the sensitivity of the algorithm is not better than RBMS.

Table 3: Algorithm Sensitivity for Divergent Sequences

| Exon    | RBMS | BLASTZ |
|---------|------|--------|
| H3 %    | 21   | 19     |
| Hsc70 % | 81   | 57     |
| PCNA %  | 60   | 55     |
| SPD %   | 38   | 33     |
| APE1 %  | 60   | 40     |

We considered another data set of Viruses and Ecoli from Genbank, there are about 50 sequences and we used BLASTZ, MUMmer and our algorithm on these to compare the quality of the sequence alignment.

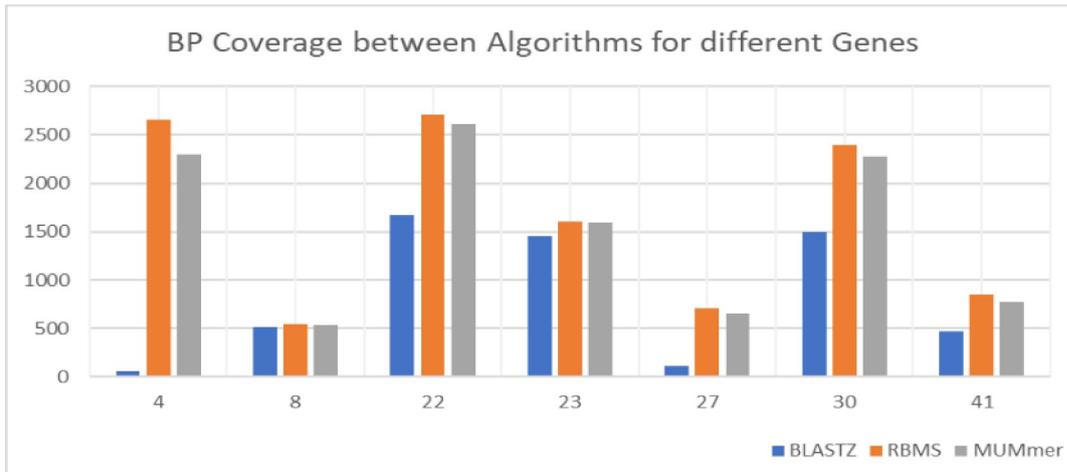

Figure 8: Comparison of BP coverage across data set sequences

In this experiment, we compare the BP coverage, a base pair coverage is where similar base. For example, a DNA sequence is made up of adenine (A), guanine (G), cytosine (C) and thymine (T). In this experiment, we compare how many Adenines in One sequences is aligned to Adenine of the other, and similarly for the other 3 bases. The experiment here is to see how many of such base pairs (BP) are perfectly aligned in the sequences. In these 50 sequences, there were some sequences which were too small to mention and hence were omitted.

To make the experiment board and whole, we considered another data set of 47 Viruses and Ecoli sequences from Genbank. The algorithms used were the same BLASTZ, MUMmer, LAlign along with our algorithm.

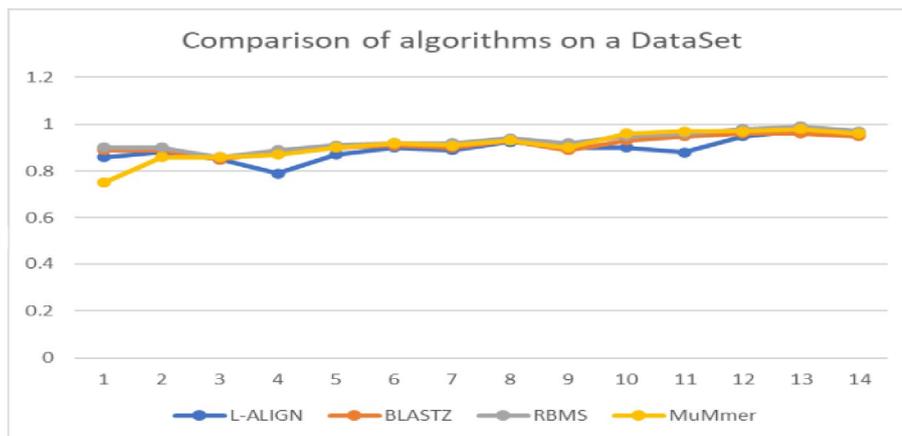

In this comparison, we used the ratio of matches to an optimal alignment algorithm like Needleman-Wulsh Alignment algorithm. 1 being a perfect match. From the experiments, we can see that the ratio across most of the algorithms is the same except L-Align which was marginally lower. RBMS performed marginally better on 7 occasions, which is attributed mainly because of the recursive step where the length of the seeds is varied until it is less than 8 characters in the inter MMSS region and perhaps due to our faster stitching algorithm in the later stages.

All results were performed on computing environment with Intel dual core machine, with 16 GB ram and windows 10 64-bit operating system. We see from the performance that our algorithm is not as fast as BLASTZ but is faster than the brute force method. In our experiments, for the brute force method, the computer hanged several times when the sequences were increased to 400000 and 500000. Our algorithm worked perfectly fine for this sequence length but was marginally slower to BLASTZ. we feel that the recursion step in our algorithm to find the perfect match seeds in the inter MMSS regions to be the main cause why it is not faster than BLASTZ [9]. These results set us nicely to expand on other AI rules, which we are currently exploring [31] and perhaps experiment on different kinds of AI rules.

## 5. CONCLUSION

In this paper we have introduced a new algorithm to find patterns using Suffix tree along with rule engine. The algorithm is fast enough but at the same time it is sensitive which is good balance to have when there is more importance placed for sensitivity than speed. For randomly generated sequences, the algorithm is faster than brute force method and slightly slower to BLASTZ version of BLAST algorithm. However, for homologous and divergent sequences, especially divergent sequences it is very sensitive than BLASTZ but is not as sensitive as the brute force method. The algorithm employs a small set of rules in the rule engine to find these patterns. We would like to grow these rules and perhaps add more intelligence in to these rules to develop and train in the future. In the future, we would like to completely find all matches with the inter MMSS region although we are not sure whether that would make any biological sense or increase sensitivity. Also, we would like to parallelize the algorithm and perhaps use this algorithm to find multiple sequence alignments.

## ACKNOWLEDGEMENTS

The authors would like to thank NIMHANS and IBM and family members for supporting us.

**Authors**

Suchindra is an Engineering from Bangalore University. He is currently pursuing neuroscience and genome research at National Institute of Mental Health and Neurosciences. His research areas are Brain haemorrhages, Autism and Genome research. He has been active researcher for 10 years. He is working for Karnataka State Govt.

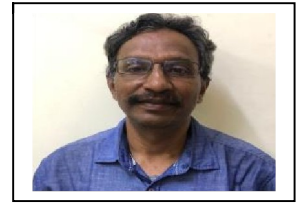

Preetham Nagaraj is a Computer Engineer from Bangalore University. He is currently pursuing Genome research, more so related to virus breakouts, causes and solutions. He is currently working at IBM Services.

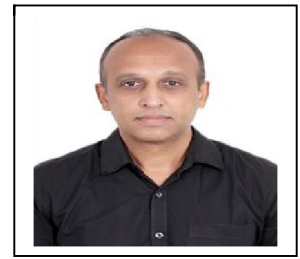